\documentclass[11pt,a4paper]{article}

\usepackage[square,sort&compress,comma,numbers]{natbib} 
\usepackage{amsmath,amsfonts,graphicx,natbib,psfrag,color,bm}
\usepackage[textsize=small]{todonotes}
\usepackage{algorithm,algpseudocode}
\usepackage{booktabs}
\usepackage{csquotes}
\usepackage{comment}
\usepackage{url}
\usepackage{textcomp}
\usepackage{lscape}
\usepackage{longtable}
\usepackage{multirow}
\usepackage{mathtools}
\usepackage{color}
\usepackage{wrapfig} 
\usepackage{tikz}
\usetikzlibrary{arrows.meta}
\usetikzlibrary{shapes,arrows}
\tikzset{%
  >={Latex[width=2mm,length=2mm]},
            base/.style = {rectangle, rounded corners, draw=black,
                           minimum width=1.3cm, minimum height=1cm,
                           text centered, font=\sffamily},
  		blue/.style = {base, fill=blue!30},
       cyan/.style = {base, fill=cyan!30},
       red/.style = {base, fill=red!30},
        orange/.style = {base, fill=orange!15},
        green/.style = {base, fill=green!30},
}
\setcitestyle{authoryear,open={(},close={)}}

\title{Bayesian inference for a spatio-temporal model of road traffic collision data}%Application of a sinusoidal dynamic linear model to road traffic collision data
\author{Nicola Hewett$^a$, Andrew Golightly$^b$, Lee Fawcett$^a$ \&  Neil Thorpe$^c$}
\date{$^a$School of Mathematics, Statistics and Physics, Newcastle University,\\
  Newcastle upon Tyne, NE1 7RU, UK\\
  $^b$Department of Mathematical Sciences, Durham University,\\
Stockton Road, Durham, DH1 3LE, UK\\
  $^c$School of Engineering, Newcastle University,\\
  Newcastle upon Tyne, NE1 7RU, UK}

\begin{document}
\maketitle
\begin{abstract}
Improving road safety is hugely important with the number of deaths on the world’s roads remaining unacceptably high; an estimated 1.35 million people die each year (WHO, 2020). Current practice for treating collision hotspots is almost always reactive: once a threshold level of collisions has been exceeded during some predetermined observation period, treatment is applied (e.g. road safety cameras). However, more recently, methodology has been developed to predict collision counts at potential hotspots in future time periods, with a view to a more proactive treatment of road safety hotspots. Dynamic linear models provide a flexible framework for predicting collisions and thus enabling such a proactive treatment. In this paper, we demonstrate how such models can be used to capture both seasonal variability and spatial dependence in time course collision rates at several locations. The model allows for within- and out-of-sample forecasting for locations which are fully observed and for locations where some data are missing. We illustrate our approach using collision rate data from 8 Traffic Administration Zones in North Florida, USA, and find that the model provides a good description of the underlying process and reasonable forecast accuracy.
\end{abstract}


\section{Introduction}
Every year the lives of approximately 1.3 million people are cut short as a result of a road traffic crash. Between 20 and 50 million more people suffer non-fatal injuries, with many incurring a disability as a result of their injury \citep{WHO}. 
%Typically, we need to wait before a number of collisions or fatalities occur before taking a precaution; an advantage of forecasting collisions is when we preempt a large number of collisions, an intervention will be put in place, and hence be proactive instead of reactive.
Working with collision counts can introduce issues of zero-inflation, especially over short time-frames. By working with rates over zones, we have the advantage of fewer zeros in the data set and upon removing these, we may treat the data as continuous, which can be mathematically convenient in terms developing a tractable model. Most road traffic data are recorded sequentially over time and it is common for there to be dependencies between each observation. Hence, it is necessary to account for these dependencies in the model via a time-series model, such as a state-space model. The use of state-space models in road safety analysis is relatively new and uncommon, though they provide advantages for prediction.

State-space models can be used for modelling univariate or multivariate time-series in the presence of non-stationarity, structural changes and irregular patterns \citep[see e.g.][]{harvey1990forecasting,west2006bayesian}. 
%For each zone in the Florida data, the rate of collisions appears fairly predictable, since it repeats quite regularly its behavior over time: seasonality. Hence we can use a  time-series model with a seasonal component.
Time-series analysis typically begins with the formulation of a model that accounts for temporal dependence, for example through auto-correlation, trend or seasonality. The use of state-space models within a time-series setting allows for uncertainty quantification in both the observation process and any dynamic variables that are not observed directly. Forecasting therefore accounts for these different sources of uncertainty and, when inferences are made within the Bayesian paradigm, additional parameter uncertainty. Throughout, we focus on a particular class of state-space model within which the observation and system equations involve linear functions of the latent process. Such models are known as dynamic linear models \citep[DLMs, see e.g.][]{west2006bayesian,petris2009dynamic} and offer several practical benefits over their nonlinear counterparts. Notably, they admit a tractable observed data likelihood function, allowing a computationally efficient approach to inference and forecasting.

\cite{gamerman1993dynamic} give a list of hierarchical dynamic linear models (DLMs) used for the state evolution, smoothing and filtering through the stages of the hierarchy. Although state-space models and DLMs in particular, have been to date rarely exploited in the road safety context \citep[see e.g.][]{fei2011bayesian,buddhavarapu2015bayesian}, they have been ubiquitously applied in environmental settings. For example, \cite{lai2020sequential} use a spatio-temporal model to forecast sensor output consisting of temperature and humidity measurements at five locations in North East England. The signal is described using coupled dynamic linear models, with spatial effects specified by a Gaussian process (GP). A related approach in the context of emissions data can be found in \cite{shaddick2002modelling}.

Our contribution is a joint spatio-temporal model of collision rates over multiple zones. A DLM is used at the level of a single zone, and allows for seasonality via a single harmonic with time-varying amplitude and phase parameters. We then account for spatial dependence at nearby locations by adding a spatial Gaussian process to the system equation, thereby smoothing spatial deviations from the underlying temporal process. The resulting model allows for both within- and out-of-sample forecasting for locations which are fully observed and for locations at which some data are missing. A Bayesian approach is used to infer both dynamic and static model components and leverages the tractability of the observed data likelihood, which can be efficiently computed via a forward filter \citep[see e.g.][]{carter1994gibbs,fruhwirth1994data}. We apply the inference scheme to a real data application consisting of monthly collision rate data from North Florida, USA within fixed Traffic Administration Zones. We assess the assumption of a time-varying parameters governing the seasonal component to each zone separately before considering a joint model of all zones.

The remainder of this paper is organised as follows. A brief description of the data is given in Section~\ref{Sec:data}. The structure of the DLM for a single zone and joint zones is given in Section~\ref{sec:DLM}. In Section~\ref{Sec:BI} we outline the details of the Bayesian inference scheme, before considering the real data application in Section \ref{sec:applic}. Conclusions are drawn in Section~\ref{sec:Conc}.

\section{Data}
\label{Sec:data}
\begin{figure}[H]
\centering
\includegraphics[width=0.49\textwidth,height=0.4\textheight]{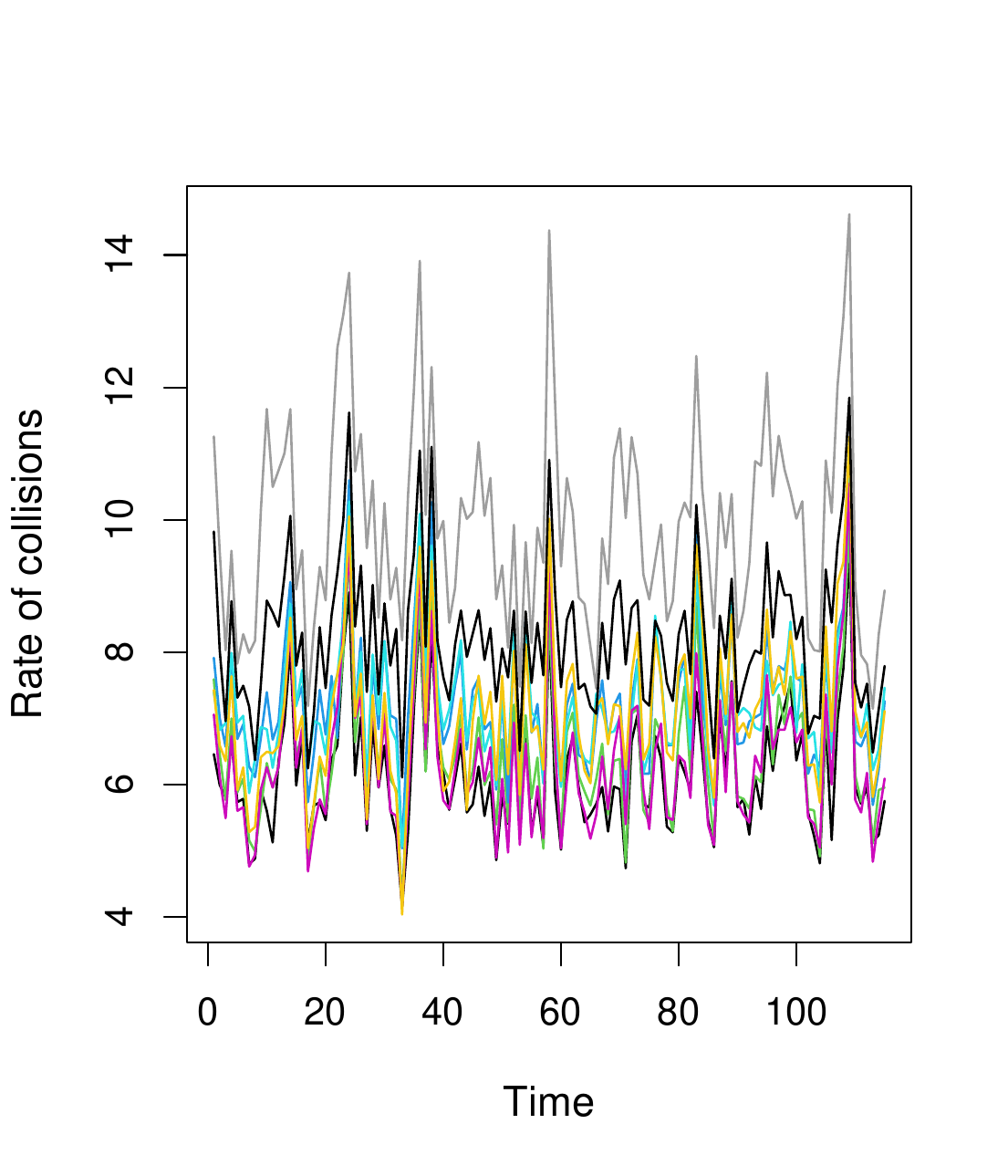}
\includegraphics[width=0.49\textwidth]{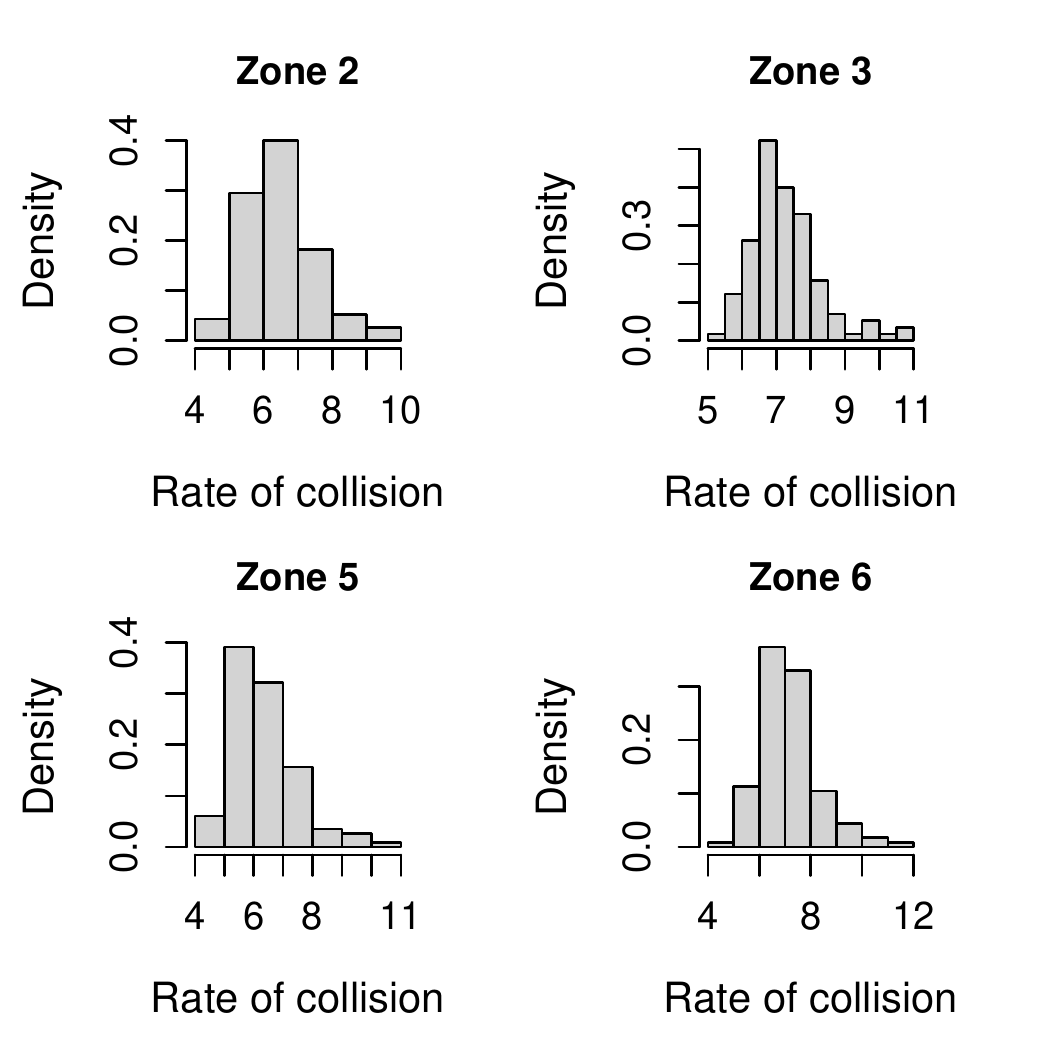}
\caption{Left: Time series plots of monthly collision rate in each of the 8 zones. Right: Histograms of collision rates for zones 2,3,5,6.}
\label{fig:TimeSeries8}
\end{figure}

\begin{figure}[H]
\centering
\includegraphics[width=0.495\textwidth]{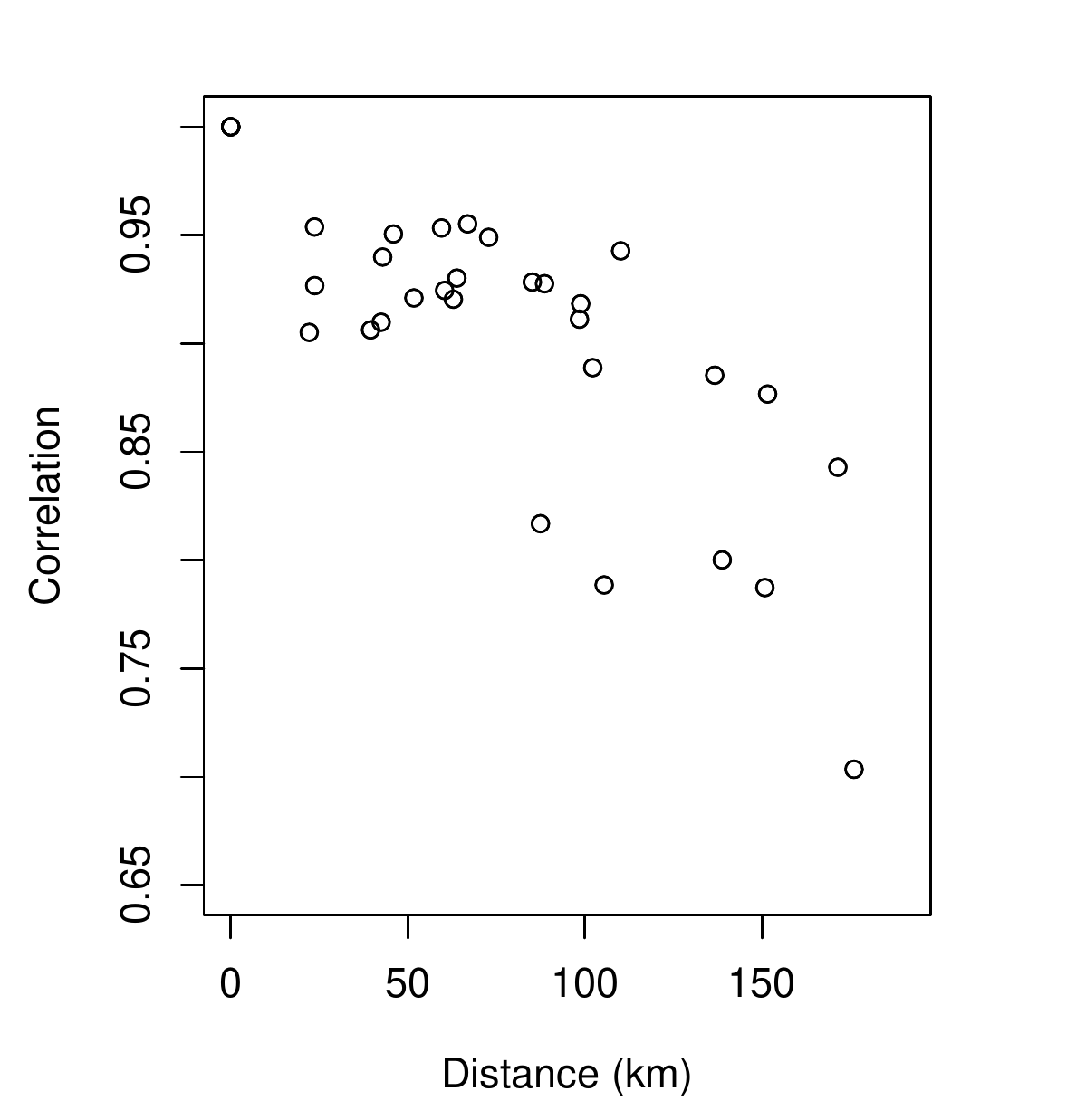}
\includegraphics[width=0.495\textwidth,height=0.37\textheight]{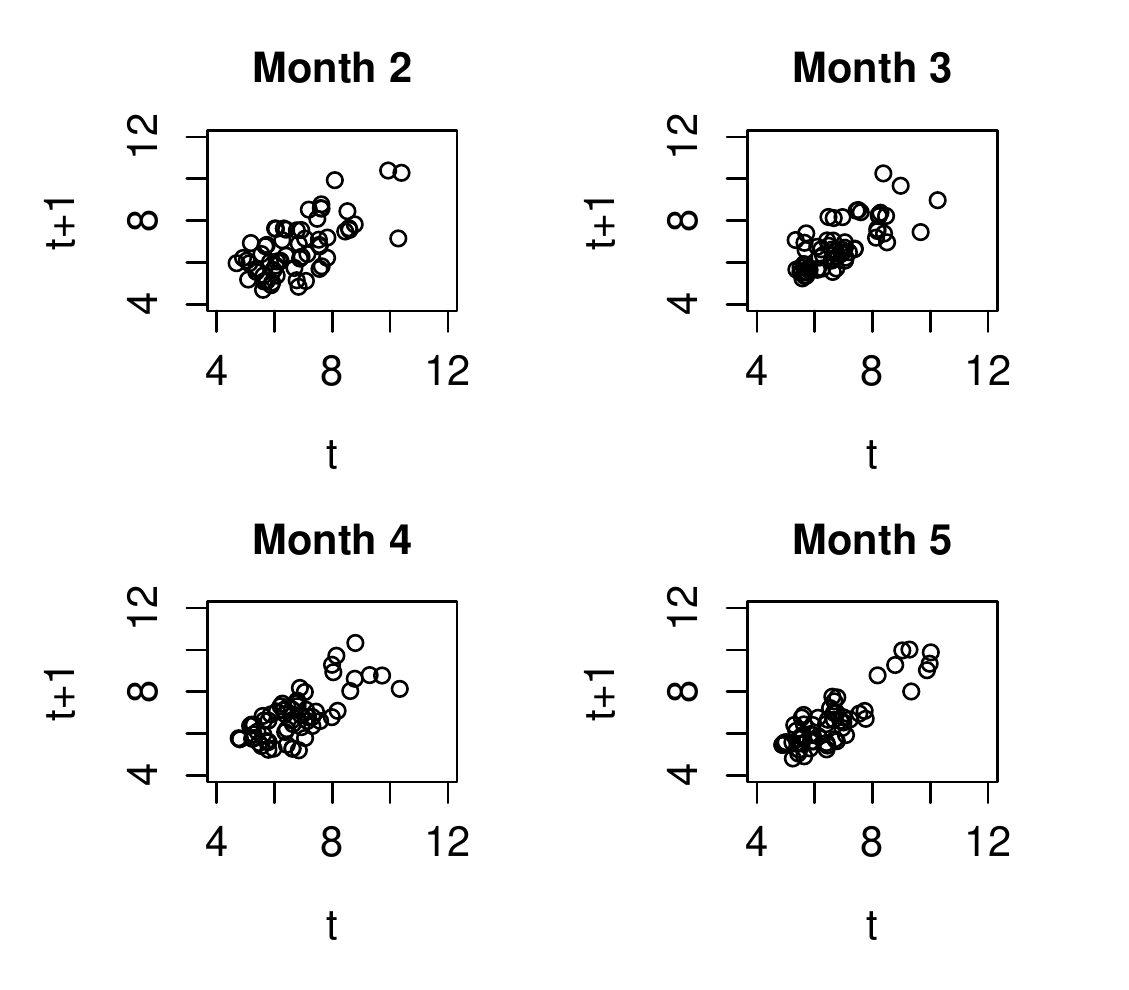}
\caption{Left: Correlation between the 8 zones against distance between zones (km). Right: The temporal dependence between observations in months 2,3,4,5 in consecutive years across all zones.}
\label{fig:Temp8Zone}
\end{figure}

We consider monthly collision rate data available from North Florida, USA within fixed Traffic Administration Zones. There are 8 zones in which collision counts have been tracked and recorded at multiple sites. The rate of collisions per zone was then calculated as the average number of collisions across those sites in each month. For each zone we have 115 months of observations where the most recent observations are from April 2014.  Figure~\ref{fig:TimeSeries8} shows the multiple data streams over time for the different zones. For all zones, the monthly collision rates exhibit sinusoidal patterns over a 12 month period. Histograms of the monthly collision rates suggest that a Gaussian observation model may adequately describe the observation process. Through scatter plots, we determined that there was clear temporal dependence between certain months in year $t$ to year $t+1$, precluding the use of a simpler model with ``month'' as a fixed effect. Furthermore, zones geographically closer are more strongly correlated (see Figure~\ref{fig:Temp8Zone}). 


\section{Dynamic linear model (DLM)}
\label{sec:DLM}
\begin{figure}[!ht]
\label{fig:statespace}
%\hspace*{-0.0cm}% shifts left by xcm
\centering
\scalebox{0.8}{
\begin{tikzpicture}[node distance=1.0cm,
    every node/.style={fill=white, font=\sffamily}]
  % Specification of nodes (position, etc.)
  \node (start)             [green]              {$\theta_{t_0}$};
  
  \node (above1)     [orange, right of=start, xshift=3cm]          {$\theta_{t_1}$};
  
 \node (above2)     [orange, right of=start, xshift=6.5cm]          {$\theta_{t_2}$};

\node (above3)     [orange, right of=start, xshift=11cm]
{$\theta_{t_{i-1}}$}; 

\node (above4)     [orange, right of=start, xshift=14.5cm]
{$\theta_{t_{i}}$}; 
  
\node (below1)     [cyan, right of=start, xshift=3cm, yshift=-2.5cm]          {$x_{t_1}$};
  
 \node (below2)     [cyan, right of=start, xshift=6.5cm, yshift=-2.5cm]          {$x_{t_2}$};

\node (below3)     [cyan, right of=start, xshift=11cm, yshift=-2.5cm]
        {$x_{t_{i-1}}$};

\node (below4)     [cyan, right of=start, xshift=14.5cm, yshift=-2.5cm]          {$x_{t_{i}}$};

% Lower text
\node (bottext1)     [right of=start, xshift=3cm,yshift=-3.5cm]          {$ $};

\node (bottext2)     [right of=start, xshift=6.5cm,yshift=-3.5cm]          {$ $};

\node (bottext3)     [right of=start, xshift=11cm,yshift=-3.5cm]          {$ $};

\node (bottext4)     [right of=start, xshift=14.5cm,yshift=-3.5cm]          {$ $};

  % Specification of lines between nodes specified above
  % with aditional nodes for description 

    % Initialisation line
    \draw[ultra thick,dotted,magenta]	(2,1) -- node[xshift=-1.2cm,yshift=-1cm] {Initialisation} (2,-3.5);	

    % Right arrows on top line
   
	\draw[->]      (start) -- node[yshift=0.5cm] {$\pi(\theta_{t_1} | \theta_{t_0})$} (above1) ;
	
	\draw[->]      (above1) -- node[yshift=0.5cm] {$\pi(\theta_{t_2} | \theta_{t_1})$} (above2) ;
	
	\draw[->]      (above2) -- node[] {$\cdots$} (above3) ;
	
	\draw[->]      (above3) -- node[yshift=0.5cm] {$\pi(\theta_{t_{i}} | \theta_{t_{i-1}})$} (above4) ;
	
%	\draw[->]      (above4) -- node[yshift=0.5cm] {\cdots} (22,0) ;
	 \draw[->]	(above4) -- node[] {$\cdots$} (18,0);
	% Downward arrows left to right
   
	\draw[->]      (above1) -- node[] {$\pi(x_{t_1} | \theta_{t_1})$} (below1) ;
	
	\draw[->]      (above2) -- node[] {$\pi(x_{t_2} | \theta_{t_2})$} (below2) ;
	
	\draw[->]      (above3) -- node[] {$\pi(x_{t_{i-1}} | \theta_{t_{i-1}})$} (below3) ;
	
	\draw[->]      (above4) -- node[] {$\pi(x_{t_{i}} | \theta_{t_{i}})$} (below4) ;
	
  \end{tikzpicture}
  }
  \caption{Directed acyclic graph showing the dependence structure of the state-space model.}
\end{figure}
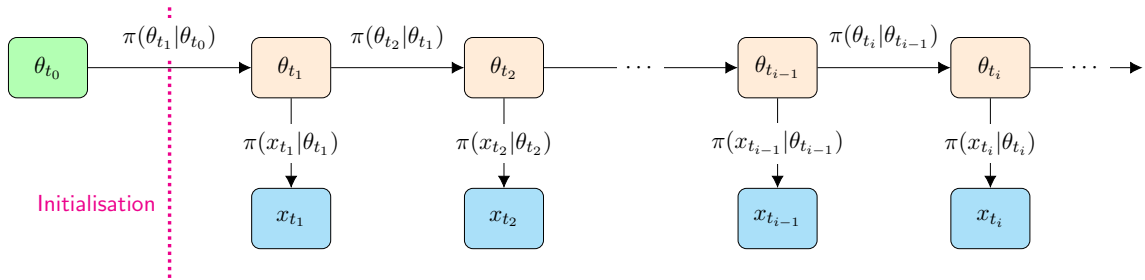

State-space models build on the relatively simple dependence structure of a (first order) Markov chain (in that information about some state $\theta_{t_i}$ carried by all previous values of the chain up to time $t_{i-1}$ is the same as that carried by $\theta_{t_{i-1}}$ alone). They are made of two main components, observed data ($x_{t_1}, \ldots, x_{t_n}$) and unobserved/latent states ($\theta_{t_0}, \ldots, \theta_{t_n}$). Figure~\ref{fig:statespace} shows the evolution of a simple univariate state-space in which the continuous valued latent state process $\{ \theta_{t_0}, \theta_{t_1}, \ldots, \theta_{t_{i-1}}, \theta_{t_i}, \ldots \}$ evolves according to a first order Markov chain with transition density $\pi(\theta_{t_i}|\theta_{t_{i-1}})$. The continuous-valued observation process $\{x_{t_1},x_{t_2},\ldots,x_{t_{i-1}},x_{t_i},\ldots\}$ is linked to the latent state process at an arbitrary time $t_i$ via the density $\pi(x_{t_i}|\theta_{t_i})$; here it is assumed that the observed data are conditionally independent given the latent states. The observable process $\{X_{t_i}\}$ depends on the underlying, unobservable latent state process $\{\theta_{t_i}\}$ and we can reasonably assume that the observation $X_{t_i}$ only depends on the state of the system at the time the measurement is taken, $\theta_{t_i}$. It remains that we specify the relationship between $X_{t_i}$ and $\theta_{t_i}$, and between $\theta_{t_i}$ and $\theta_{t_{i-1}}$. In each case, we adopt linear relationships, and further assume that the errors in the state and observed components are independent and normally distributed. This structure leads to a dynamic linear model (DLM), given by the following equations:
\begin{equation*}
\begin{aligned}
\text{Observation Equation}: \hspace{0.5cm} X_{t_i} & = F_{t_i}\theta_{t_i} +\nu_{t_i} \\
\text{System Equation}: \hspace{0.5cm} \theta_{t_i} & = G_{t_i}\theta_{t_{i-1}} + \omega_{t_i} \\
\end{aligned}
\end{equation*}
Here, $X_{t_i}$ is a scalar, $\theta_{t_i}$ is a $p \times 1$ vector, $F_{t_i}$ is a $1 \times p$ vector, $G_{t_i}$ is a $p \times p$ matrix and $\nu_{t_i} \sim$ N$(0, V_{t_i})$ and $\omega_{t_i} \sim $N$(0, W_{t_i})$ are independent white noise processes with known variance matrices $V_t$ and $W_t$, typically assumed to be constant. Assuming that the initial latent state follows a Gaussian distribution gives
\begin{equation*}
\begin{aligned}
\theta_0 & \sim  \text{N}(m_0, C_0) \\
\theta_{t_i}|\theta_{t_{i-1}} & \sim \text{N}(G_{t_i} \theta_{t_{i-1}}, W_{t_i})  \\
X_{t_i}|\theta_{t_i} & \sim \text{N}(F_{t_i}\theta_{t_i}, V_{t_i}) \\
\end{aligned}
\end{equation*}
 for suitably chosen hyperparameters $m_0$ and $C_0$. In what follows we consider a DLM appropriate for data at a single zone, before considering a joint model over all zones.

\subsection{Zone specific model}

The data set described in Section~\ref{Sec:data} showed seasonality in that, over all zones there was a clear sinusoidal pattern about the rate of collisions over a year. Therefore, to account for this within the DLM we include a single harmonic. Note that it is possible to account for seasonality through the inclusion of multiple harmonics in the system equation \citep[see e.g.][]{petris2009dynamic}, however, we find that using a single harmonic and allowing the amplitude and phase to vary over time, provides a parsimonious modelling approach. 

Consider first a single location. We assume constant variance matrices $V$ and $W$ and data at irregularly spaced times $t_1,t_2, \ldots, t_n$. The observation equation is
\begin{equation}
X_{t_i}=F_{t_i}\theta_{t_i}+\nu_{t_i}, \hspace{0.5cm} \nu_{t_i} \stackrel{indep}{\sim} \text{N}(0,V),
\label{eq:ObsEq}
\end{equation}
where $\theta_{t_i}=(\theta_{1,t_i},\theta_{2,t_i},\theta_{3,t_i})^T$ and the observation matrix is given by
\begin{equation*}
F_{t_i}=\left(\sin\left(\frac{2\pi i}{P_x}\right),\cos\left(\frac{2\pi i}{P_x}\right),1\right)
\end{equation*}
where $P_x$ is the time corresponding to one complete period ($P_x=12$ for seasonal data).
Note that the observation equation can be written as 
\begin{equation}
X_{t_i}=\tilde{\theta}_{1,t_i}\cos\left(\frac{2\pi i}{P_x} - \tilde{\theta}_{2,t_i}\right)+\theta_{3,t_i}+\nu_{t_i}
\label{eq:tildeParams}
\end{equation}
where the dynamic parameters in Eq.~(\ref{eq:ObsEq}) and (\ref{eq:tildeParams}) are related using
\begin{equation}
\label{trans}
\tilde{\theta}_{1,t_i}=\sqrt{\theta_{1,t_i}^2+\theta_{2,t_i}^2}, \hspace{1cm} \tilde{\theta}_{2,t_i}=\tan^{-1}\left(\frac{\theta_{1,t_i}}{\theta_{2,t_i}}\right).
\end{equation}
We impose some smoothness in these dynamic parameters by taking the system equation to be of the form
\begin{equation*}
    \theta_{t_i} = \theta_{ t_{i-1}} + k_{t_i}\omega_{t_i}, \hspace{0.5cm} \omega_{t_i} \stackrel{indep}{\sim} \text{N}(0, W)
\end{equation*}
which has been further altered to allow for measurements that are irregularly spaced on a temporal grid. That is, we include a coefficient, $k_{t_i}$, in the variance in the state equation such that $k_{t_i}^2=t_i-t_{i-1}$. Hence, the sinusoidal form DLM captures seasonality via a single harmonic whose amplitude $\tilde{\theta}_{1,t_i}$ and phase $\tilde{\theta}_{2,t_i}$ vary according to two transformed independent random walk processes.

\subsection{Joint model over zones}\label{jm}
We now consider a model of monthly collision rates that captures both the seasonality, and additionally, the correlation between nearby zones. 
Let $X_{t_i}=(X_{t_i}^1, \ldots, X^{n_z}_{t_i})^T$ denote the collection of monthly collision rates at time $t_i$ with $X_{t_i}^j$ corresponding to zone $j$, and $j=1,\ldots,n_z$. In Section~\ref{sec:singleZone} we find that amplitude and phase are plausibly constant for each zone. Therefore, for ease of notation, we fix $\theta_{1,t_i}=\theta_1$ and $\theta_{2,t_i}=\theta_2$ in what follows. The model at zone $j$ is
\begin{equation*}
\label{eq:multsite}
\begin{aligned}
X_{t_i}^{j}&=\theta_1^{j}\sin\frac{\pi t_i}{6}+\theta_2^{j}\cos\frac{\pi t_i}{6}+\theta_{3,t_i}^{j}+ \nu_{t_i}^{j}, \hspace{0.5cm} \nu_{t_i}^{j} \stackrel{indep}{\sim} N(0,V^{j}),\\
\theta_{3,t_i}^{j}&=\theta_{3,t_{i-1}}^{j}+k_{t_i}\omega_{t_i}^{j}+p_{t_i}^{j}, \hspace{0.5cm} \omega_{t_i}^{j} \stackrel{indep}{\sim} N(0,W^{j}), \\
\end{aligned}
\end{equation*}
To induce correlation between nearby zones, we include the term $p^j_{t_i}$ as a component of a spatially smooth error process $p_{t_i}= (p_{t_i}^1, \ldots, p_{t_i}^{n_z})^T$. We model $\{p_{t_i}$, $t_i = 1, \ldots, n\}$ using independent (over $i$) zero-mean Gaussian processes so that 
\begin{equation*}
p_{t_i} \stackrel{indep}{\sim} GP\{0,f_3(\cdot; \eta_3)\}.
\end{equation*}  
We impose smoothness by taking a squared exponential kernel for the covariance function. Hence, the covariance between spatial errors at locations $j$ and $j'$ is
\begin{equation}
f_3(d_{jj'}; \eta_3)=\text{Cov}(\theta_{3,t_i}^{j},\theta_{3,t_i}^{j'})=\sigma_3^2\text{exp}(-\phi_3 d_{jj'}),
\label{eq:covariance}
\end{equation}
with $\eta_3=(\sigma_3, \phi_3)^T$ parameterising the kernel; note that $\phi_3$ determines the decay ratio of the correlation as the distance between sites $j$ and $j'$ $(d_{jj'})$ increases (\cite{banerjee2014hierarchical}). Similarly, we adopt GP priors for $\theta_1=(\theta_1^1, \ldots, \theta_1^{n_z})^T$ and $\theta_2=(\theta_2^1, \ldots, \theta_2^{n_z})^T$ so that $\theta_1 \sim GP(m_1(\cdot), f_1(\cdot ; \eta_1))$ and $\theta_2 \sim GP(m_2(\cdot), f_2(\cdot ; \eta_2))$ with $f_1$ and $f_2$ defined analogously to Eq.~(\ref{eq:covariance}) with the addition of $m_1(\cdot)$ and $m_2(\cdot)$ as appropriate mean functions. Hence, the full spatial DLM model (over all locations) is 
\begin{equation*}
    \begin{aligned}
    X_{t_i}=& F_{t_i}\theta_{t_i}+\nu_{t_i}, \hspace{0.5cm} \nu_{t_i} \stackrel{indep}{\sim} \text{N}(0,\text{diag}\{V^1, \ldots, V^{n_z}\}),\\
    \theta_{3, t_i} =& \theta_{3, t_{i-1}} + k_{t_i}\omega_{t_i}, \hspace{0.5cm} \omega_{t_i} \stackrel{indep}{\sim} \text{N}(0, \text{diag}\{W^1, \ldots, W^{n_z}\}+K_3),
    \end{aligned}
\end{equation*}
where $F_{t_i}= \text{diag}(F_{t_1}^1, \ldots, F_{t_i}^{n_z})$, $\theta_{3,t_i}= (\theta_{3, t_i}^1, \ldots, \theta_{3,t_i}^{n_z})^T$, $\theta_{t_i}=(\theta_1^1,\theta_2^1,\theta_{3,t_i}^1, \ldots, \theta_1^{n_z},\theta_2^{n_z},\theta_{3,t_i}^{n_z})^T$ and $K_3$ is an $n_z \times n_z$ matrix with $(i,j)th$ element $f_3(d_{ij},\eta_3)$.

\section{Bayesian inference}
\label{Sec:BI}
For simplicity, suppose we have $n_z$ zones with $n$ observations in each zone. Let $V=(V^1, \ldots, V^{n_z})^T$ and $W=(W^1, W^2, \ldots, W^{n_z})^T$. Furthermore, let $\eta_3=(\sigma_3, \phi_3)^T$ denote the hyperparameters governing $f_3(\cdot)$, with $\eta_1=(\sigma_1, \phi_1)^T$ and $\eta_2=(\sigma_2, \phi_2)^T$ denoting the hyperparameters governing $f_1(\cdot)$ and $f_2(\cdot)$ respectively. Let $x^j= (x_{t_1}^j, \ldots, x_{t_n}^j)^T$ denote the vector of collision rates at site $j$ so that $x=(x^1,\ldots, x^{n_z})$ denotes the complete data set over all zones. The joint posterior over all dynamic and static parameters is proportional to the marginal static parameter posterior multiplied by the conditional posterior of the dynamic process $\theta_3=(\theta_{3,t_{0}},\ldots,\theta_{3,t_{n}})$ such that
\begin{equation*}
    \begin{aligned}
    \pi(\theta_1, \theta_2, V, W, &\eta_1, \eta_2, \eta_3, \theta_3 | x) \propto \\
     &\pi(\theta_1, \theta_2, V, W, \eta_1, \eta_2, \eta_3 | x) \times \pi(\theta_3 | \theta_1, \theta_2, V, W, \eta_1, \eta_2, \eta_3, x).   
    \end{aligned}
\end{equation*}
Let $\psi$ denote all fixed model parameters. To simulate realisations from the joint posterior we use a two step approach:
\begin{enumerate}
    \item Simulate from the marginal posterior $\psi \sim \pi(\psi | x)$.
    \item Simulate from the conditional posterior $\theta_3 \sim \pi(\theta_3 | \psi, x)$.
\end{enumerate}
For step 1, as the marginal static parameter posterior is intractable, we use Markov chain Monte Carlo \citep[see e.g.][]{gilks1995markov}. For step 2 we use a forward filter backward sampling algorithm 
\citep[see e.g.][]{west2006bayesian} to directly draw from $\pi(\theta_3|\psi,x)$. We provide details as follows.

\subsection{Simulation based inference}
 Let $\theta_{3,t_{0:n}}=(\theta_{3,t_0},\theta_{3,t_1},\ldots,\theta_{3,t_n})$ denote the collection of latent states up to time $t_n$ and let $x=x_{t_{1:n}}=(x_{t_1},\ldots,x_{t_n})$ denote the observed data. Note that $\theta_{3,t_i}=(\theta_{3,t_i}^1,\ldots, \theta_{3,t_i}^{n_z})^T$ and $x_{t_i}=(x_{t_i}^1,\ldots,x_{t_i}^{n_z})^T$. Upon assuming an independent prior specification for the constituent terms of $\psi$, Bayesian inference may proceed as follows.
 Integrating out the dynamic parameters, gives us the marginal posterior:
\begin{equation*}
\begin{aligned}
\pi(\theta_1, \theta_2, &V, W, \eta_1, \eta_2, \eta_3|x) \propto \\
& \pi(\theta_1| \eta_1)\pi(\theta_2| \eta_2) \left[ \prod_{j=1}^{n_z} \pi(V^j)\pi(W^j)\right]\times\\
& \pi(\eta_1)\pi(\eta_2)\pi(\eta_3) \times \pi(x| \theta_1, \theta_2, \eta_3, V, W)\\
\end{aligned}
\end{equation*}
where the marginal likelihood $\pi(x|\theta_1,\theta_2,\eta_3,V,W)$ is given by
\begin{equation}
\pi(x|\theta_1,\theta_2,\eta_3,V,W)=\pi(x_{t_1}|\theta_1,\theta_2,\eta_3,V,W)\prod_{i=1}^{n-1} \pi(x_{t_{i+1}}|x_{t_{1:i}},\theta_1,\theta_2,\eta_3,V,W)
\label{eq:MargLike}
\end{equation}
and whose constituent terms are analytically tractable. Moreover, $\pi(\theta_1|\eta_1)=N(\theta_1; m_1,K_1)$ and $\pi(\theta_2|\eta_2)=N(\theta_2; m_2,K_2)$ are multivariate normal densities, $\pi(V^j)$ and $\pi(W^j)$ are the prior densities ascribed to $V^j$ and $W^j$, $\pi(\eta_1)$, $\pi(\eta_2)$ and $\pi(\eta_3)$ are the prior densities ascribed to $\eta_1$, $\eta_2$ and $\eta_3$. 

The marginal likelihood can be efficiently evaluated using a forward filter. It will be helpful here to define
\begin{equation*}
\tilde{X}_{t_i} \equiv X_{t_i}-  \theta_1\sin\frac{\pi t_i}{6}-\theta_2\cos\frac{\pi t_i}{6}=\tilde{F}_{t_i}\theta_{3,t_i}+ \nu_{t_i}, \hspace{0.5cm} \nu_{t_i} \stackrel{indep}{\sim} \text{N}(0,\text{diag}\{V^1, \ldots, V^{n_z}\}),\\
\end{equation*}
so that
\begin{equation*}
    \tilde{X}_{t_i}|\theta_{3,t_i} \sim N(\tilde{F}_{t_i}\theta_{3,t_i},\textrm{diag}\{V\}),
\end{equation*}
where $\tilde{F}_{t_i}$ is the $n_z \times n_z$ identity matrix and will be omitted for ease of notation in what follows. We also write
\[
\theta_{3,t_i}|\theta_{3,t_{i-1}} \sim N(\theta_{3,t_{i-1}}, \tilde{W}_{t_i})
\]
where $\tilde{W}_{t_i}=k_{t_i}^2 (\text{diag}\{W^1, \ldots, W^{n_z}\}+K_3)$.

\begin{algorithm}[H]
\caption{Forward filter}
\begin{algorithmic}[1]
	\item[] 
	\item[1.] Initial distribution: $\theta_{3,t_0} \sim N(m_0, C_0)$. Store the values of $m_0$ and $C_0$.
	\item[2.] For $t_i, i=1,\ldots,n,$
	\begin{enumerate}
		\item[(a)] Prior at $t_i$. Using the system equation, we have that
		\begin{equation*}
		\theta_{3}|\tilde{x}_{t_{1:i-1}} \sim N( m_{t_{i-1}}, C_{t_{i-1}} + \tilde{W}_{t_i}).
		\end{equation*}
		Store $R_{t_i}= C_{t_{i-1}} + \tilde{W}_{t_i}.$ 
		\item[(b)] One step forecast. Using the observation equation, we have that
		\begin{equation*}
		\tilde{X}_{t_i}|\tilde{x}_{t_{1:i-1}} \sim N(m_{t_{i-1}}, \ R_{t_i}+V).
		\end{equation*}
		Store the marginal likelihood contribution
		\begin{equation*}
		\pi(\tilde{x}_{t_i}|\tilde{x}_{t_{1:i-1}})=N(\tilde{x}_{t_i}; m_{t_{i-1}},  R_{t_i} + V).
		\end{equation*}
		\item[(c)] Posterior at $t_i$: $\theta_{3,t_i}|\tilde{x}_{t_{1:i}} \sim N(m_{t_i}, C_{t_i})$ where
		\begin{equation*}
		\begin{aligned}
		m_{t_i} &= m_{t_{i-1}} + R_{t_i}( R_{t_i} +V)^{-1}(\tilde{x}_{t_i} -m_{t_{i-1}}),\\
		C_{t_i}&= R_{t_i} - A_{t_i}Q_{t_i}A_{t_i}^T,
		\end{aligned}
		\end{equation*}
		where $A_{t_i}=R_{t_i}Q_{t_i}^{-1}$ and $Q_{t_i}=R_{t_i} +V$.
		Store the values of $m_{t_i}$ and $C_{t_i}$.
\end{enumerate}
\end{algorithmic}
\label{alg:FF}
\end{algorithm}
Algorithm~\ref{alg:FF} gives the steps of the forward filter. We see that  the constituent terms in Eq.~(\ref{eq:MargLike}) are obtained from the forward pass as
\begin{equation*}
\pi(\tilde{x}_{t_i}|\tilde{x}_{t_{1:i-1}},\theta_1,\theta_2,\eta_3,V,W)=\text{N}(\tilde{x}_{t_i}; m_{t_{i-1}},  R_{t_i} + V),
\end{equation*}
where %$\psi=(V,W, \theta_1, \theta_2, \textcolor{pink}{\mathbf{\sigma, \phi}})$,
$R_{t_{i}}=C_{t_{i-1}} + \tilde{W}_{t_i}$ and $m_{t_{i-1}}, C_{t_{i-1}}$ are updated recursively; we refer the reader to \cite{petris2009dynamic} \citep[see also][]{carter1994gibbs,fruhwirth1994data,west2006bayesian} for further details.

Although the marginal likelihood is tractable, the posterior will typically be unavailable in closed form. Hence we use Metropolis-Hastings to generate draws from $\pi(\psi|\tilde{x})$; see Algorithm~\ref{alg:MCMC}.

\begin{algorithm}[H]
\caption{MCMC scheme}
\begin{algorithmic}[1]
	\item[1] Initialise the chain with $\psi^{(0)}$. Set $r=1$.
	\item[2] Propose $\psi^{*} \sim q(\psi^{*}|\psi^{(r-1)})$.
	\item[3] Calculate the acceptance probability $\alpha(\psi^{*}|\psi^{(r-1)})$ of the proposed move, where
	\begin{equation*}
	\begin{aligned}
	\alpha(\psi^{*}|\psi^{(r-1)})&=\text{min}\left\{1,A(\psi^{*}|\psi^{(r-1)})\right\}\\
	&= \text{min}\left\{1, \frac{\pi(\psi^{*}|\tilde{x}_{1:n})q(\psi^{(r-1)}|\psi^{*})}{\pi(\psi^{(r-1)}|\tilde{x}_{1:n})q(\psi^{*}|\psi^{(r-1)})} \right\}\\
	\end{aligned}
	\end{equation*}
	\item[4] With probability $\alpha(\psi^{*}|\psi^{(r-1)})$, set $\psi^{(r)}=\psi^{*}$; otherwise set $\psi^{(r)}=\psi^{(r-1)}$.
	\item[5] Set $r:=r+1$. Return to step 2.

\end{algorithmic}
\label{alg:MCMC}
\end{algorithm}
It remains that, given draws of $\psi^{(1)}, \ldots, \psi^{(N)}$ we can sample $\theta_3^{(r)} \sim \pi(\theta_3 | \psi, x)$, $r=1,\ldots, N$. This can be achieved by noting the factorisation 
\begin{equation*}
    \pi(\theta_3|\psi,x) = \pi(\theta_{3,t_n}|\psi,x_{t_{1:n}})\prod_{i=0}^{n-1} \pi(\theta_{3,t_i}| \theta_{3,t_{i+1}}, \psi, x_{t_{1:i}})
\end{equation*}
where the constituent densities are tractable and can be sampled recursively via a backward sampling algorithm. The key steps are given in Algorithm~\ref{alg:BackSamp}.

\begin{algorithm}[H]
\caption{Backward sampler}
\begin{algorithmic}[1]

	%\item[] \textbf{Backward Sampling}
	\item[3.] Sample $\theta_{3,n}|\tilde{x}_{t_{1:n}} \sim N(m_n, C_n).$
	\item[4.] For $t_i, i=n,\ldots,1$,
	\begin{enumerate}
		\item[(a)] Backward distribution: $\theta_{3,t_i}|\theta_{3,t_{i+1}},\tilde{x}_{t_{1:i}} \sim N(h_{t_i}, H_{t_i})$, where
		\begin{equation*}
		\begin{aligned}
		h_{t_i} &= m_{t_i} + C_{t_i}(C_{t_i} + \tilde{W}_{t_{i+1}})^{-1}(\theta_{3,t_{i+1}}-m_{t_{i}}),\\
		H_{t_i} &= C_{t_i} - C_{t_i}^T(C_{t_i} + \tilde{W}_{t_{i+1}})^{-1}C_{t_i}.
		\end{aligned}
		\end{equation*} 
		\item[(b)] Sample $\theta_{3,t_i}|\theta_{3,t_{i+1}},\tilde{x}_{t_{1:i}} \sim N(h_{t_i}, H_{t_i}).$
	\end{enumerate}
\end{algorithmic}
\label{alg:BackSamp}
\end{algorithm}

\subsubsection*{Missing data}
Missing observations are commonplace, that is, only observations on a subset of components of $X_t$ may be available at time $t_i$. To account for this in the model we let $\tilde{X}_{t_i}^o$ denote the observed rates at time $t_i$. The observation model is then written as 
\begin{equation}
\tilde{X}_{t_i}^o=P_{t_i}\tilde{X}_{t_i}
\end{equation}
 where the $n_{obs} \times n_z$ incidence matrix $P_{t_i}$ determines which components are observed at time $t_i$ (\cite{lai2020sequential}). For example, if we have data from 5 zones and data are missing at the second and third zone at time $t_i$, then the incidence matrix is
  \begin{equation*}
  P_{t_i}=
  \begin{pmatrix}
  1 & 0 & 0 &0 &0 \\
  0 & 0 & 0 &1 &0 \\
  0 & 0 & 0 &0 &1 \\
  \end{pmatrix}. 
  \end{equation*} 
The forward filter and backward sampler can be modified straightforwardly to allow for this scenario. In brief, each occurrence of $\tilde{F}_{t_i}$ is replaced by $P_{t_i}\tilde{F}_{t_i}$ and each occurrence of $V$ is replaced by $P_{t_i}V$ in Algorithm~1.

\subsection{Within-sample predictive density}
\label{sec:WSP}
In order to assess model fit, we consider the within-sample predictive density. The within-sample predictive density is given by 
\begin{equation*}
\pi(\hat{x}_{t_{1:n}}|x_{t_{1:n}})=\int\int \pi(\hat{x}_{t_{1:n}}|\theta_{3,t_{1:n}},\psi)\pi(\theta_{3,t_{1:n}},\psi|x_{t_{1:n}}) d\theta_{3,t_{1:n}}d\psi
\end{equation*}
where
\begin{equation*}
\pi(\theta_{3,t_{1:n}},\psi|x_{t_{1:n}})=\pi(\theta_{3,t_{1:n}}|\psi,x_{t_{1:n}})\pi(\psi|x_{t_{1:n}}).
\end{equation*}
Although the within-sample predictive density is intractable, draws from $\pi(\theta_{3,t_{1:n}},\psi|x_{t_{1:n}})$ are readily available and therefore $\pi(\hat{x}_{t_{1:n}}|x_{t_{1:n}})$ can be obtained via Monte Carlo. Given draws $(\psi^{(r)},\theta_{3,t_{1:n}}^{(r)}), r=1,\ldots,N$, we can simulate 
\begin{equation}\label{pred}
\hat{X}_{t_i}^{(r),j}|\theta_{t_i}^{(r),j},\psi^{(r),j} \sim N(F_{t_i}\theta_{t_i}^{(r),j},V^{(r),j}), \hspace{0.5cm} r=1,\ldots,N,\hspace{0.2cm} i=1,\ldots,n,\hspace{0.2cm} j=1,\ldots,n_z,
\end{equation}
where $\theta_{t_i}^{(r),j}=(\theta_1^{(r),j}, \theta_2^{(r),j}, \theta_{3,t_i}^{(r),j})$ denotes the $r$th sample of $\theta_{t_i}^j$, with $\hat{X}_{t_i}^{(r),j}$ defined similarly. Draws obtained from (\ref{pred}) can be summarised (e.g. via the mean, upper and lower quantiles) and bench-marked against the observed data. 

\subsection{$k$-step ahead prediction}
\label{sec:Kstep}
The system and observation forecast distributions can be obtained by exploiting the linear Gaussian structure of the DLM. The one-step ahead system forecast density is given by
\begin{equation*}
\begin{aligned}
\pi(\theta_{3,t_{n+1}}|x_{t_{1:n}})&=\int \int \pi(\theta_{3,t_{n+1}}| \theta_{3,t_n}, \psi, x_{t_{1:n}})\pi(\theta_{3,t_n}| \psi, x_{t_{1:n}})\pi(\psi|x_{t_{1:n}}) d\theta_{3,t_n} d\psi \\
&=\int \pi(\theta_{3,t_{n+1}}|\psi,x_{t_{1:n}})\pi(\psi|x_{t_{1:n}})d\psi
\end{aligned}
\end{equation*}
where
\begin{equation*}
\pi(\theta_{3,t_{n+1}}|\psi, x_{t_{1:n}}) = N(\theta_{3,t_{n+1}}; m_{t_n}, C_{t_n} +\tilde{W}_{t_{n+1}}). 
\end{equation*}
Similarly, the one-step ahead observation forecast density is given by
\begin{equation*}
\pi(x_{t_{n+1}}|x_{t_{1:n}}) = \int \pi(x_{t_{n+1}}|\psi, x_{t_{1:n}})\pi(\psi|x_{t_{1:n}}) d\psi
\end{equation*}
where
\begin{equation*}
\pi(x_{t_{n+1}}|\psi, x_{t_{1:n}}) = N(x_{t_{n+1}}; m_{t_n}, C_{t_n} + \tilde{W}_{t_{n+1}} + V).
\end{equation*}
Hence, given $N$ posterior summaries $(m_{t_n}^{(r)},C_{t_n}^{(r)}),r = 1,...,N$ from $\pi(\theta_{3,t_n} |\psi,x_{t_{1:n}})$ and $\psi^{(r)}$ from $\pi(\psi|x_{t_{1:n}})$, the one-step ahead state and observation forecast distributions can be sampled via Monte Carlo, by drawing
\begin{equation*}
\begin{aligned}
\theta_{3,t_{n+1}}^{(r)}| \psi^{(r)}, x_{t_{1:n}} &\sim N(m_{t_n}^{(r)}, C_{t_n}^{(r)} + \tilde{W}^{(r)}_{t_{n+1}}),\\
\tilde{X}_{t_{n+1}}^{(r)}| \psi^{(r)}, x_{t_{1:n}} &\sim N(m_{t_n}^{(r)}, C_{t_n}^{(r)} + \tilde{W}^{(r)}_{t_{n+1}} + V^{(r)}).
\end{aligned}
\end{equation*}
Then, $X_{t_{n+1}}^{(r)}$ can be obtained from $\tilde{X}_{t_{n+1}}^{(r)}$ by adding the term $\theta_1^{(r)}\sin\frac{\pi t_{n+1}}{6}+\theta_2^{(r)}\cos\frac{\pi t_{n+1}}{6}$ to the latter. For the general $k$-step ahead forecast, the above draws are replaced by 
\begin{equation*}
\begin{aligned}
\theta_{n+k}^{(r)}| \psi, x_{t_{1:n}} &\sim N\left\{m_{t_n}^{(r)}, R_{t_{n+k}}^{(r)} \right\},\\
\tilde{X}_{t_{n+k}}^{(r)}| \psi^{(r)}, x_{t_{1:n}} &\sim N\left\{m_{t_n}^{(r)}, R_{t_{n+k}}^{(r)} +V^{(r)}\right\},
\end{aligned}
\end{equation*}
where
\begin{equation*}
R_{t_{n+k}}^{(r)} = C_{t_n}^{(r)} + \sum_{i=1}^{k}\tilde{W}^{(r)}_{t_{n+i}}. 
%\sum_{j=0}^{k-2}W^{(j)}+W^{(j)}.
\end{equation*}
\section{Application}
\label{sec:applic}
In what follows, and where required, we implement the MCMC scheme from Section~\ref{Sec:BI} by taking a random walk proposal with Gaussian innovations. We have that $q(\psi^*|\psi)=N(\psi^*;\psi,\Sigma)$ where the innovation matrix $\Sigma = \gamma \widehat{Var}(\psi|x)$, with $\widehat{Var}(\psi|x)$ obtained from a pilot run and $\gamma$ is chosen to give an acceptance rate of around 25\% \citep{roberts2001optimal}. Within the MCMC scheme, for mathematical convenience, we will work with precisions so that $\tau_V = 1/V, \tau_{W}=1/W$. Moreover, for parameter vectors whose components must be strictly positive (i.e. $V,W, \eta$) we implement the proposal on the log scale.

\subsection{Single zone analysis}
\label{sec:singleZone}
In this section we assess the assumption that amplitude and phase vary with time. We present results for zone 4 and note similar findings (namely that amplitude and phase are plausibly constant) for the remaining zones. 

 For the single zone model, $\psi=(\tau_V, \tau_{W_1}, \tau_{W_2}, \tau_{W_3})'$ is the vector of precision parameters. We set the mean and variance of $\theta_{t_0}$ to be $m_0=(1.5, 1.5, 6)$ and $C_0=\textrm{diag}\{1.5, 1.5, 20\}$ respectively. We take an uninformative and independent prior specification for the components of $\psi$, via $\tau_V,\tau_{W_1},\tau_{W_2}, \tau_{W_3} \sim Ga(0.1, 0.1)$. The MCMC scheme was run for 22k iterations with the first 2k iterations discarded as burn-in, leaving 20k iterations on which to base posterior summaries.

\begin{figure}[H]
    \centering
    \includegraphics[]{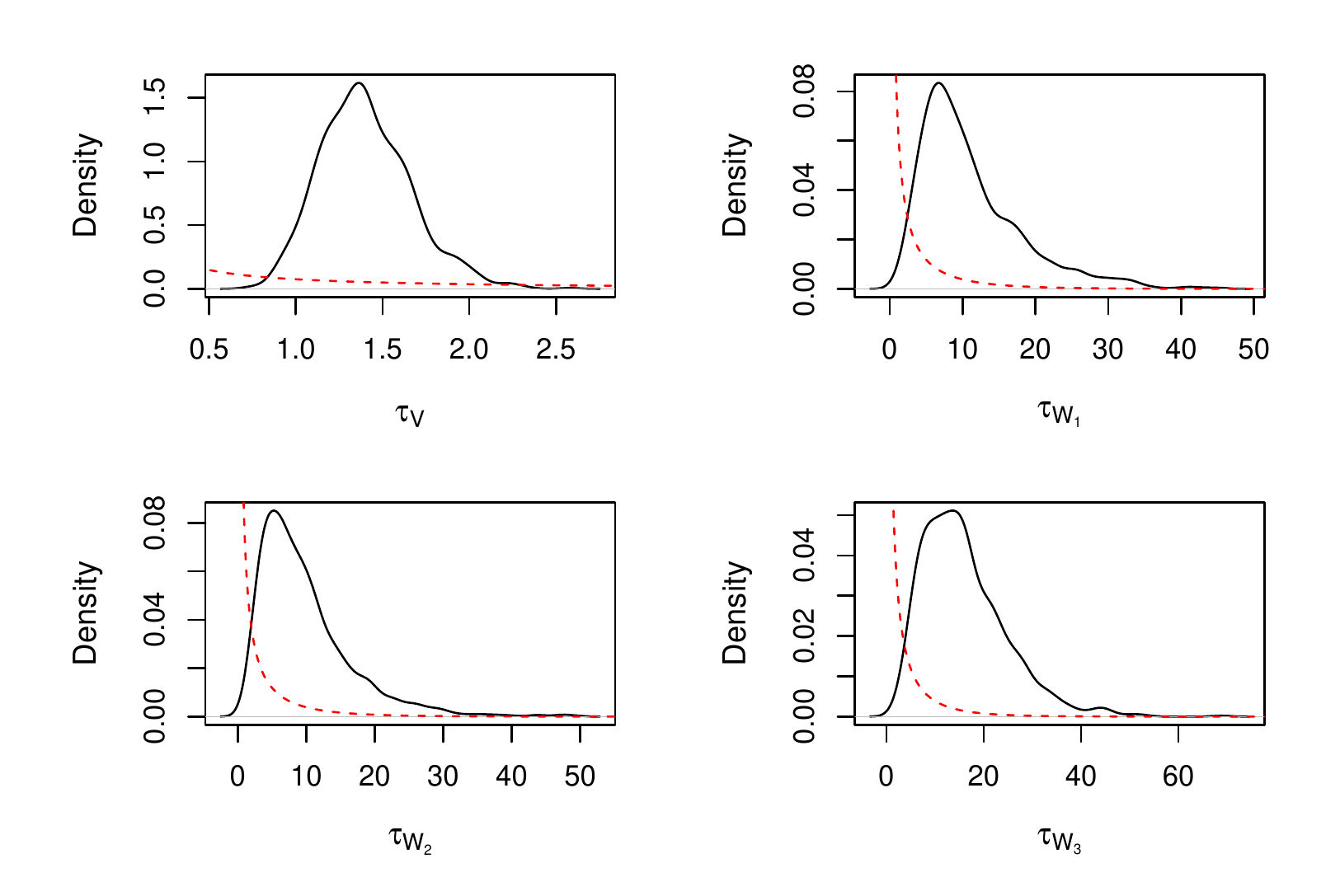}
    \caption{Density plots of $\tau_V, \tau_{W_1}, \tau_{W_2}, \tau_{W_3}$ respectively, from 20k iterations and a thin of 20 with prior densities overlaid in red. }
    \label{fig:SingleTrace}
\end{figure}

\begin{figure}[H]
    \centering
    \includegraphics[]{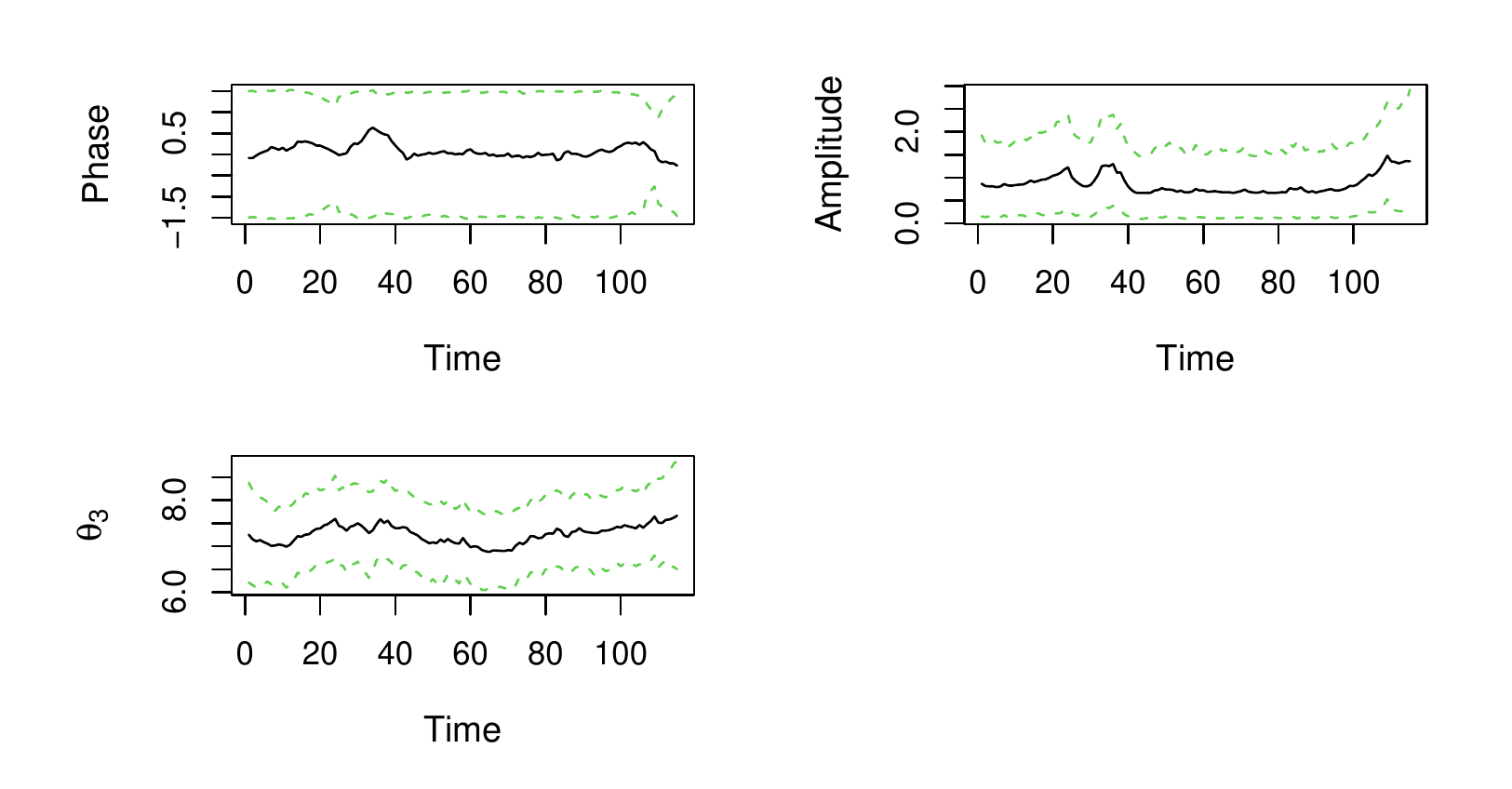}
    \caption{Phase, amplitude and $\theta_{3,t_i}$ mean and 95\% CI at zone 4 from time $t_1$ to $t_{115}$.}
    \label{fig:PhaseAmp}
\end{figure}

The marginal MH scheme gives the estimated marginal posterior densities for the components of $\psi$ shown in Figure~\ref{fig:SingleTrace} with their prior densities overlaid. 
%The analysis has been informative, resulting in marginal posterior distributions that are very different to the uninformative priors. Numerical summaries are shown in Table~\ref{tab:singlezone}. 
The $\psi$ samples were thinned to obtain 1k (near uncorrelated) draws form the marginal parameter posterior, denoted $\{\psi^{(r)}\}_{r=1}^{1000}$. The FFBS algorithm was then executed for each $\psi^{(r)}$, to obtain samples of the dynamic parameter vector, 
$\{\theta_{t_i}^{(r)}\}_{r=1}^{1000}$, $i=1,\ldots,n$, from the within-sample predictive. Samples of the dynamic components $\theta_{1,t_i}$ and $\theta_{2,t_i}$ can be transformed via (\ref{trans}) to obtain phase and amplitude draws from their respective within-sample predictive densities; see Section~\ref{sec:WSP} for further details regarding the method for obtaining samples from these predictive distributions. These distributions are summarised in Figure~\ref{fig:PhaseAmp} via their means and 95\% credible intervals. We can conclude that, upon allowing for the uncertainty in amplitude and phase, they are are plausibly constant over time for this zone. Performing the analysis on the remaining zones shows that the same conclusions can be drawn. This suggests that the dynamic parameters  $\theta_{1,t_i}$ and $\theta_{2,t_i}$, $i=1,\ldots,n$, can reasonably be replaced with static parameters $\theta_1$ and $\theta_2$. %Figure~\ref{fig:PhaseAmp} shows that $\theta_{3,t_i}$ captures the majority of the variance shown in the data and represents the overall mean. 

\begin{figure}[H]
    \centering
    \includegraphics[width=\textwidth]{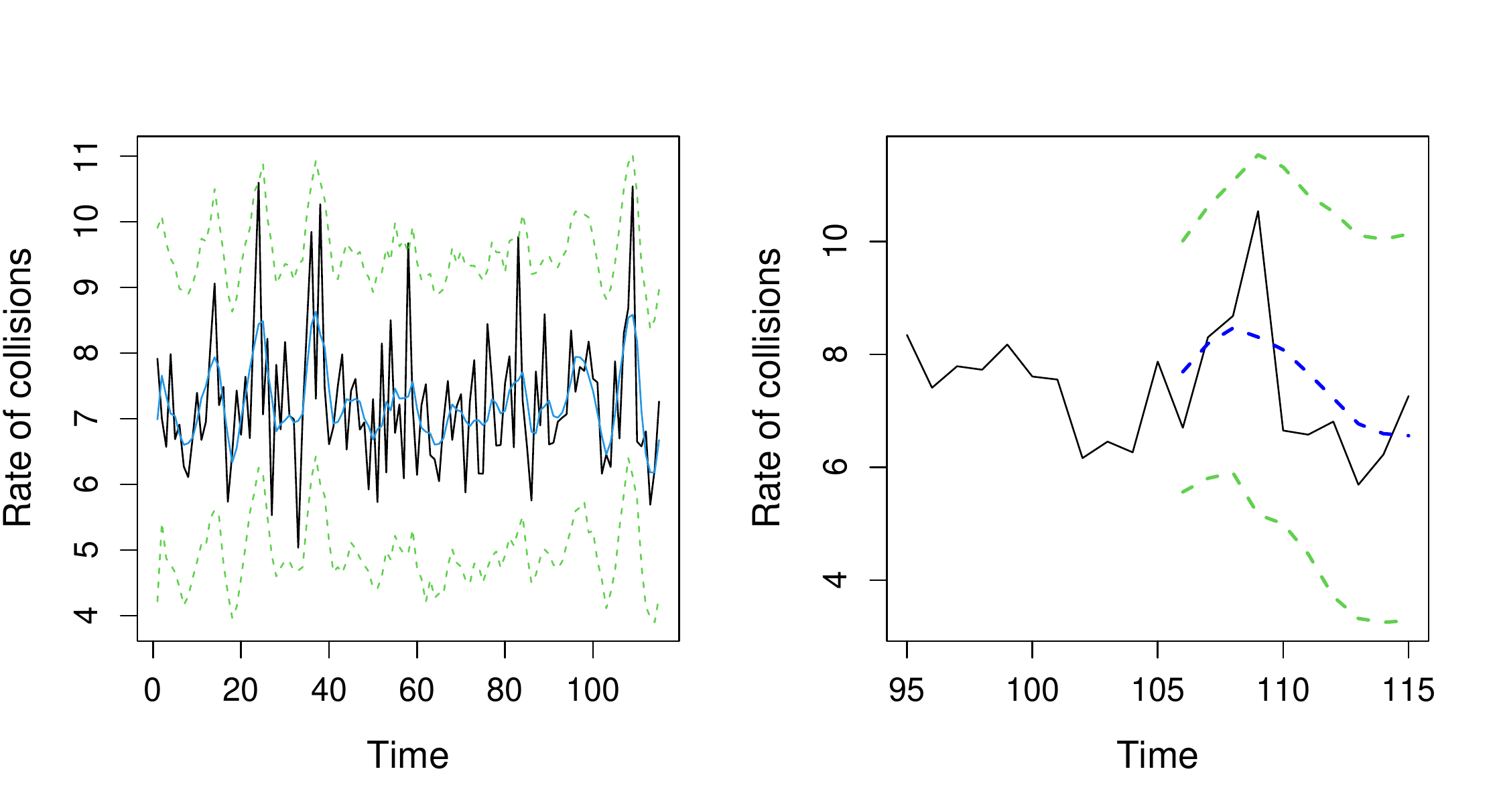}
    %\includegraphics[]{Figures/Zone4WithinOutPred.pdf}
    %\includegraphics[width=0.45\textwidth,height=6cm]{Figures/Pred10Paper.pdf}
   % \caption{Left: Zone 4 observed data (black) with overlaid within-sample predictions - mean (blue) and 95\% confidence intervals (green). Right: Observed data (black) with overlaid 10 step ahead predictions - mean (blue) with 95\% confidence interval.}
   \caption{Left: Zone 4 observed data (black) with overlaid within-sample predictions - mean (blue) and 95\% credible intervals (green). Right: Observed data (black) with overlaid 10 step ahead predictions - mean (blue) with 95\% credible interval (green).}
    \label{fig:Prediction}
\end{figure}

We assess the validity of the proposed model for a single zone by comparing observed data with their model-based within-sample posterior predictive distributions and with model-based out-of-sample forecast distributions. For the latter, we withheld the last 10 observations when fitting the model. Figure~\ref{fig:Prediction} shows the within-sample predictive distribution for the observation process, summarised by the mean and 95\% credible interval calculated for each time point. This suggests that the model is able to reasonably account for the observation process. Similarly, the 10-step ahead forecast distribution is summarised by the mean and 95\% credible interval at each time point. We see that the forecast distribution is able to capture the general trend exhibited by the observations.


\subsection{Joint zone analysis}
We now consider the joint model over all zones detailed in Section~\ref{jm}. Our prior specification takes the following form. 

We expect that amplitude and phase should be similar at nearby zones. Recall that $\theta_1 \sim GP(m_1(\cdot),  f_1(\cdot;\eta_1))$, 
$\theta_2 \sim GP(m_2(\cdot), f_2(\cdot;\eta_2))$ and the Gaussian process components in the dynamic mean process are $p_{t_i} \stackrel{indep}{\sim} GP\{0,f_3(\cdot; \eta_3)\}$ . We take the mean functions to be constant so that $m_1(\cdot)=m_2(\cdot)=1.5\mathbf{1}$, with $\mathbf{1}$ defined as an $n_z \times 1$ vector of 1s. We have that $f_k(d_{jj'}; \eta_k)=\sigma_k^2\text{exp}(-\phi_k d_{jj'})$, $k=1,2,3$. We take $\log\sigma_k \stackrel{indep}{\sim} N(\log(0.1),0.1)$ representing fairly strong prior beliefs about the amplitude variance and phase within a zone. For the logarithm of the inverse length scales, we take $\log\phi_k \sim N(\log(0.1),0.1)$ giving typical length scales of around 10km, reflecting typical distances between zones. The precisions of the observation equations governing each zone are $\tau_{V^{j}} \stackrel{indep}{\sim} Ga(0.1, 0.1)$ and similarly for the system variances, $\tau_{W^{j}} \stackrel{indep}{\sim} Ga(0.1, 0.1)$, $j=1,\ldots,n_z$. Finally, the initial values $\theta_{3,t_{0}}^j$ for each site were assumed to follow $N(6,20)$ distributions.

\begin{figure}[H]
	\centering
	\includegraphics[]{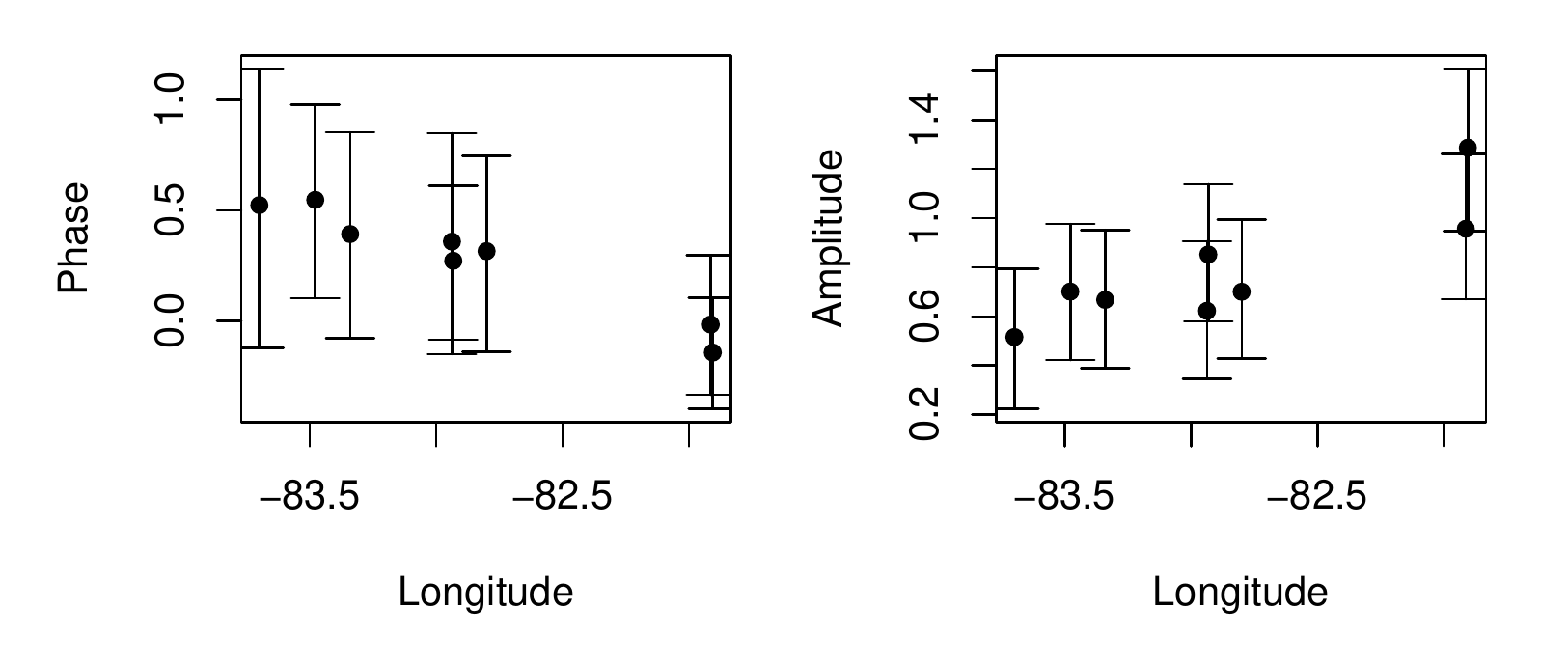}
	\caption{Mean amplitude and phase with 95\% credible intervals against longitude for each zone.}
	\label{fig:PASpatial}
\end{figure}

\begin{figure}[H]
	\centering
	\includegraphics[height=0.3\textheight]{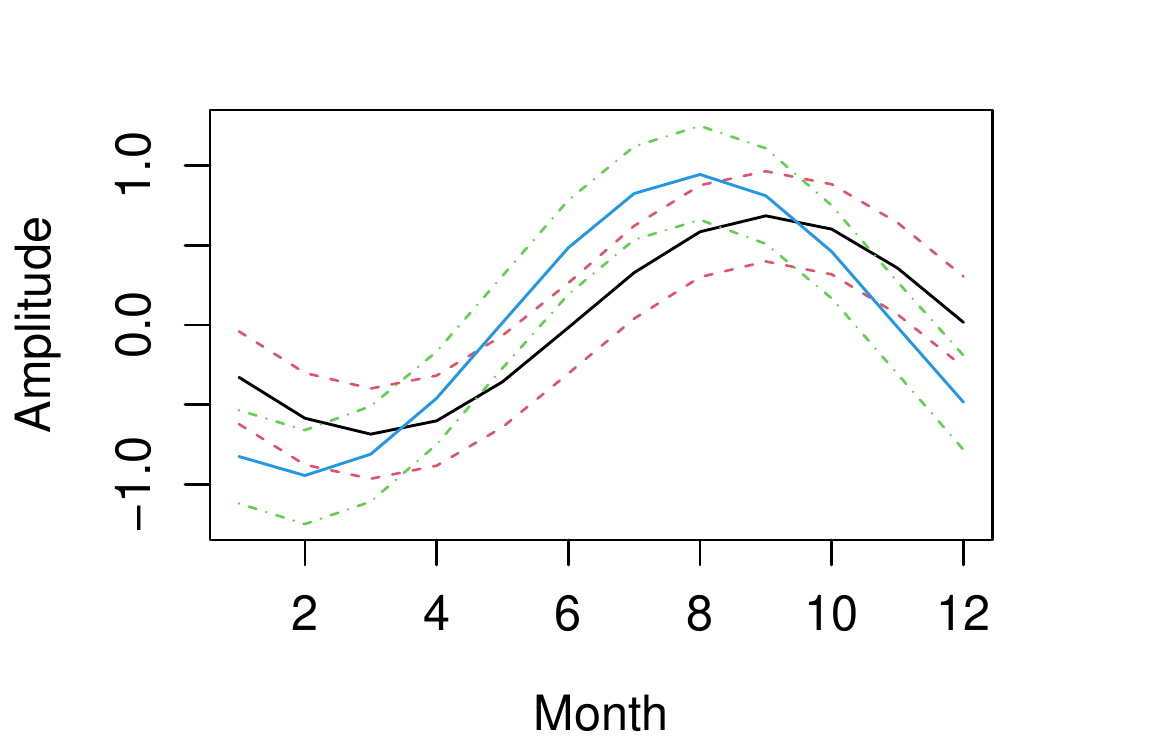}
	\caption{Mean and 95\% credible intervals for the seasonal component for the most western zone (black) against the most eastern (blue).}
	\label{fig:SineWave}
\end{figure}

The MCMC scheme was run for $1\times 10^6$ iterations; the output is summarised by Table~\ref{tab:8siteTable}. %\textcolor{red}{Figure \ref{fig:PApost} shows the marginal posterior densities for phase and amplitude at each zone - MAYBE REMOVE ALL MARGINAL DENSITY PLOTS?}, 
Figure~\ref{fig:PASpatial} shows the mean value and 95\% credible interval of the posterior densities for amplitude and phase at each zone against  longitude. There are signs of spatial dependence as the phase seems to decrease and amplitude increases in zones further to the east. Figure~\ref{fig:SineWave} shows a single period of the sine curve, averaged over draws of amplitude and phase for the most eastern versus most western zone with 95\% credible intervals. From this we would expect to see more pronounced fluctuations in the rate of collisions across the year for eastern zones. Furthermore, we would expect the highest rate of collisions to be a month sooner (August) in eastern zones than that in western zones (September). 

\begin{figure}[H]
	\centering
	\includegraphics[]{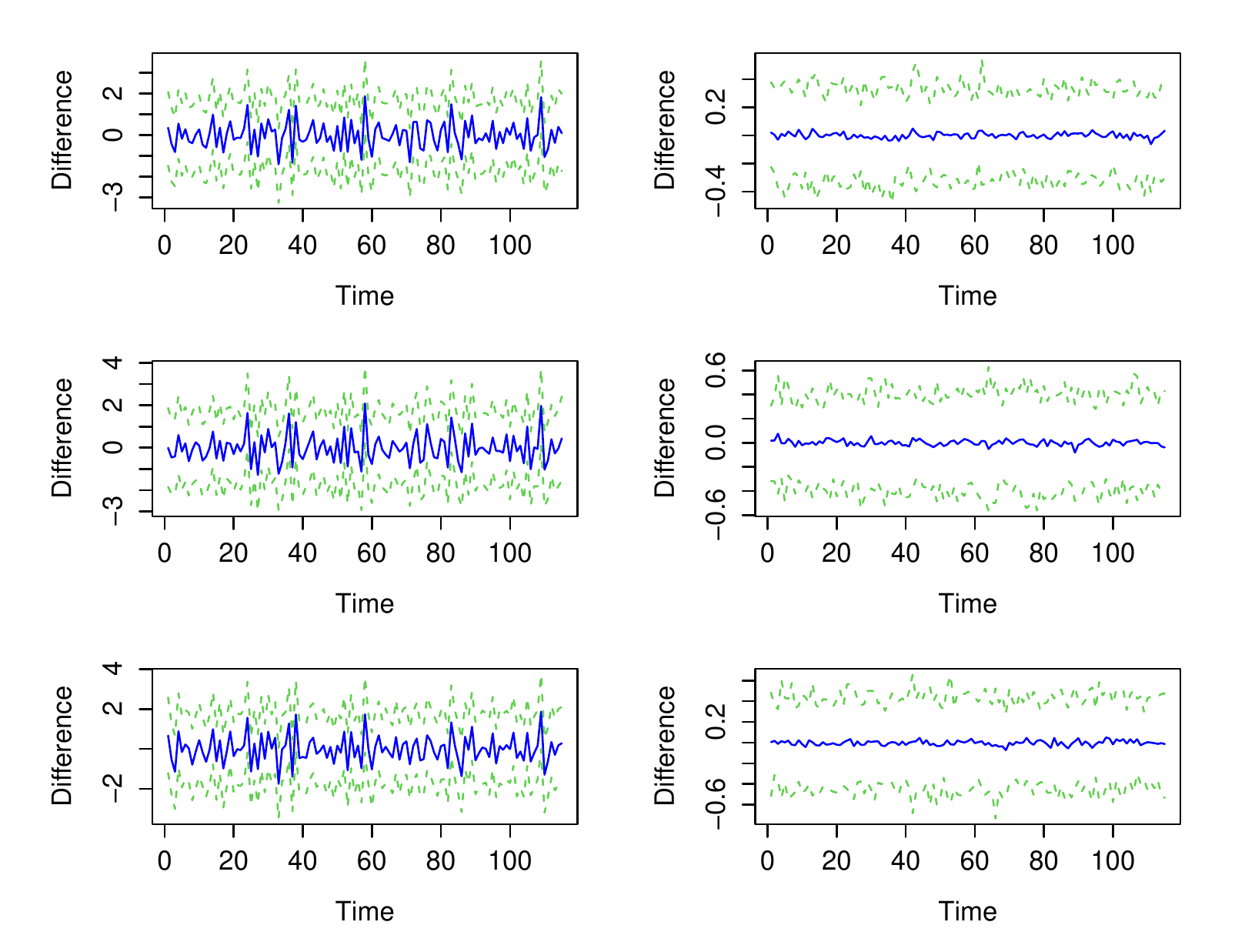}
	\caption{Mean (blue) and 95\% credible intervals (green) for the difference between the within-sample predictive and the observations over time. Each row shows the differences from the single zone analysis (left) and the joint zone analysis (right) for zones 2 (top), 4 (middle) and 8 (bottom). }
	\label{fig:Diff248}
\end{figure}

Figure~\ref{fig:Diff248} shows summaries (mean and 95\% credible interval) 
 of the difference between observations and the within-sample predicted observation process for zones 2, 4 and 8. The left-hand-side plots show the differences from the single zone analysis and the right-hand-side from the joint analysis. It is clear that the mean difference at each time-zone combination is small and that a mean difference of zero is plausible (the 95\% credible intervals include zero). Comparing left to right, shows the improvement in the within-sample predictions from a single zone analysis to a joint model; that is, the spatial information included through the GP has increased prediction precision. We additionally calculated the root mean square error (RMSE) at each time-point (observation vs prediction) and averaged this measure over all time points for each zone; the results are shown in Table~\ref{tab:RMSE}. We see that the mean RMSEs are approximately 5 times larger for the single zone analysis, giving further evidence of an improvement in fit when considering a joint model over all zones.
 
\begin{figure}[H]
	\centering
	\includegraphics[width=\textwidth]{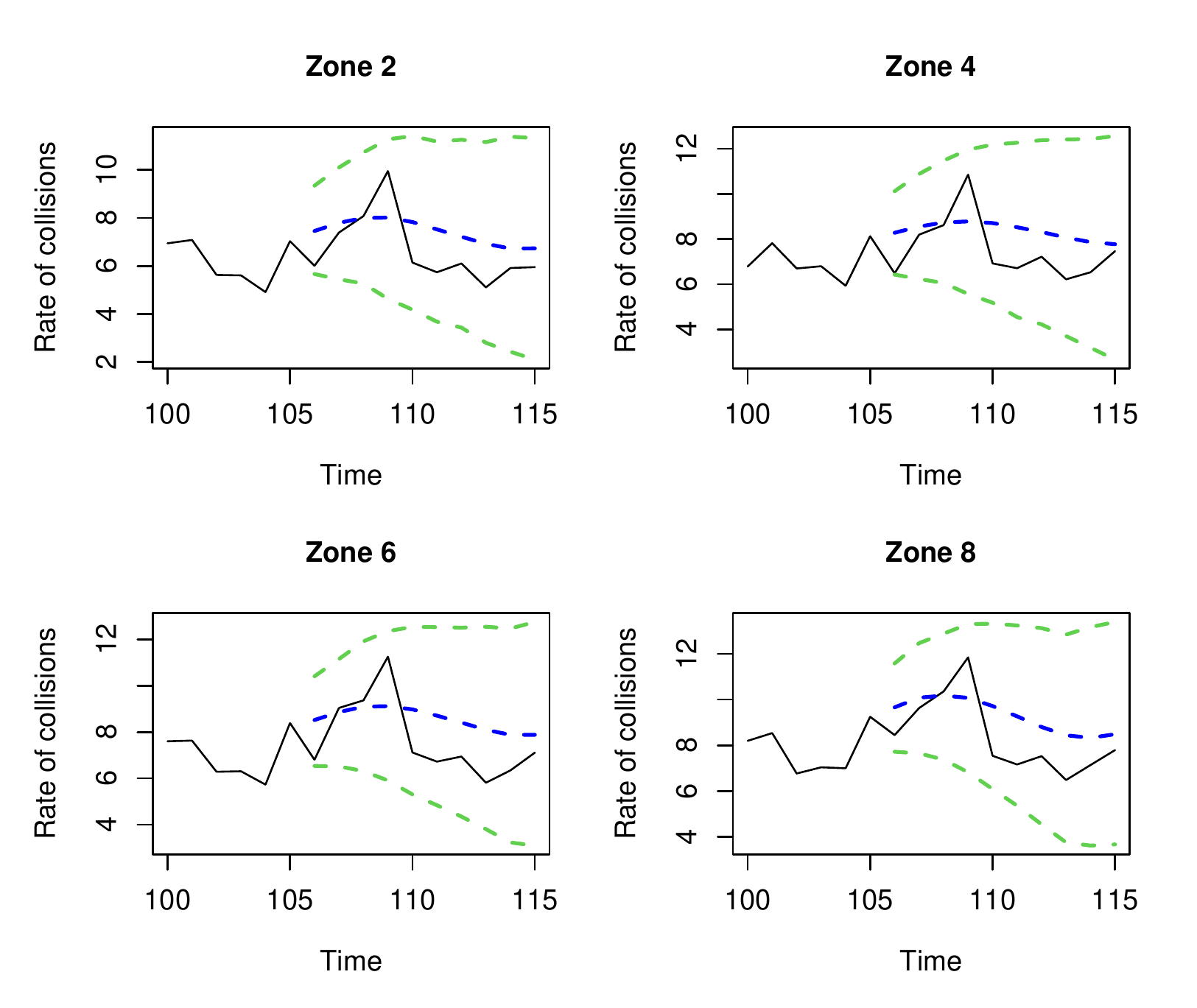}
	\caption{Rates of collisions in Florida zones 2, 4, 6 and 8 with overlaid out-of-sample 10-step ahead predictions -mean (blue) and 95\% credible intervals (green).}
	\label{fig:Florida4zone10stepPred}
\end{figure}

 Figure~\ref{fig:Florida4zone10stepPred} shows 10-step ahead predictions for zones 2, 4, 6 and 8, following application of the method in Section~\ref{sec:Kstep}. Note that the last 10 observations were removed from each zone before running the inference scheme. The figure shows that the forecast distributions are consistent with the data as they lie within the forecast intervals for all zones. As we would expect, uncertainty grows as we move away from the last recorded observation.

\begin{table}[H]
\centering
\begin{tabular}{llllllll}
\hline
$\psi$       & Mean       & 95\% CI     & $\psi$       & Mean        & 95\% CI     \\ \hline
$V^1$        & 0.034 & (0.021, 0.052) & $\theta_1^4$ & 0.251   & (-0.045, 0.532) \\
$V^2$        & 0.025 & (0.015, 0.039) & $\theta_1^5$ & 0.226   & (-0.065, 0.514) \\
$V^3$        & 0.059 & (0.039, 0.084) & $\theta_1^6$ & 0.249   & (-0.039, 0.526)  \\
$V^4$        & 0.037 & (0.022, 0.058) & $\theta_1^7$ & -0.181  & (-0.501, 0.133)  \\
$V^5$        & 0.031 & (0.018, 0.048) & $\theta_1^8$ & -0.014 & (-0.308, 0.271)  \\
$V^6$        & 0.041 & (0.023, 0.066) & $\theta_2^1$ & 0.585   & (0.301, 0.877)  \\
$V^7$        & 0.119  & (0.059, 0.196) & $\theta_2^2$ & 0.651   & (0.367, 0.944)  \\
$V^8$        & 0.045     & (0.026, 0.071)     & $\theta_2^3$ & 0.566      & (0.285, 0.856)     \\
$W^1$      & 0.021     & (0.011, 0.037)     & $\theta_2^4$ & 0.424      & (0.144, 0.722)     \\
$W^2$      & 0.024     & (0.012, 0.041)     & $\theta_2^5$ & 0.809      & (0.530, 1.098)     \\
$W^3$      & 0.023     & (0.011, 0.044)     & $\theta_2^6$ & 0.601      & (0.311, 0.896)     \\
$W^4$      & 0.025     & (0.012, 0.044)     & $\theta_2^7$ & 1.264      & (0.931, 1.587)     \\
$W^5$      & 0.024     & (0.013, 0.043)     & $\theta_2^8$ & 0.945      & (0.660, 1.249)     \\
$W^6$      & 0.034     & (0.016, 0.061)     & $\sigma_1$   & 1.688      & (1.315, 2.309)     \\
$W^7$      & 0.099     & (0.031, 0.213)     & $\sigma_2$   & 1.545      & (1.201, 2.253)     \\
$W^8$      & 0.029     & (0.014, 0.055)     & $\sigma_3$   & 1.352      & (1.349, 1.355)     \\
$\theta_1^1$ & 0.357     & (0.066, 0.642)     & $\phi_1$     & 1.527      & (1.278, 1.917)     \\
$\theta_1^2$ & 0.213    & (-0.077, 0.494)     & $\phi_2$     & 1.603      & (1.387, 1.903)     \\
$\theta_1^3$ & 0.213     & (-0.084, 0.499)     & $\phi_3$     & 1.103      & (1.098, 1.107)     \\ \hline
\end{tabular}
\caption{Marginal parameter posterior means and quantile-based 95\% credible intervals obtained from the MCMC scheme.}
\label{tab:8siteTable}
\end{table}

\begin{table}[H]
\centering
\begin{tabular}{ccc}
\hline
     & \multicolumn{2}{c}{Mean RMSE} \\
Zone & Single zone     & Joint zone     \\ \hline
1    & 1.384           & 0.198          \\
2    & 1.066           & 0.167          \\
3    & 1.214           & 0.258          \\
4    & 1.083           & 0.206          \\
5    & 1.114           & 0.185          \\
6    & 1.191           & 0.217          \\
7    & 1.339           & 0.378          \\
8    & 1.107           & 0.227          \\ \hline
\end{tabular}
\caption{The mean RMSE over all time-points for each zone, from the single zone and joint zone analyses. }
\label{tab:RMSE}
\end{table}

\section{Discussion and limitations}
\label{sec:Conc}
We have developed a spatio-temporal model for collision rates that allows for serial dependence, seasonality and correlation between rates at nearby zones. We considered a dynamic linear model (DLM) whose observation equation takes the form of a single harmonic with a smoothly time-varying amplitude and phase, thus accounting for seasonality and potential long term changes. Spatial consistency is accounted for at nearby zones by adding a Gaussian process (GP) component in the system equation. The model can be fitted in Bayesian paradigm using an efficient two-stage Markov chain Monte Carlo procedure, targeting the joint posterior over the parameters, the latent time-varying harmonic coefficients (amplitude and phase) and dynamic mean. At the first stage, parameter samples are generated from the marginal parameter posterior using a random walk Metropolis algorithm with the likelihood evaluated via a forward filter. At the second stage, samples of the dynamic parameters are generated conditionally on the static parameter draws from stage one using a backward sampler. Further details of this forward filter, backward sampling (FFBS) approach can be found in \cite{petris2009dynamic} \citep[see also][]{carter1994gibbs,fruhwirth1994data}.   

We applied our approach to a real data set consisting of 115 months of collision rates over eight traffic administration zones in Florida, USA. An exploratory analysis that considered separate models for each zone found that the phase and amplitude were plausibly constant. We were therefore able to simplify the joint model over all zones by treating the harmonic components as static, with a GP prior allowing correlation between these parameters at nearby zones. The validity of both the single zone and joint models was assessed using within-sample posterior predictive distributions, which suggested a satisfactory fit in both cases. Moreover, the within-sample predictions were improved substantially when using the joint model, with the credible intervals of our predictions narrowing almost tenfold, and a reduction in root mean squared error (RMSE) between the observations and predictions of around a factor of 5. 

Our analysis suggests clear spatial patterns between phase and longitude and amplitude and longitude. For all zones we found that the lowest rates of collisions would fall earlier in the year. The model also suggests that for western zones, the lowest rates would be in March, and in February for eastern zones. It appears that peak collision rates are in September in the East and August in the West. We would also expect to see a larger fluctuation in the rate of collisions in an eastern zone. Our interest also lies in the ability to forecast collision rates in future months. Model-based out-of-sample forecast distributions suggest that our model is able to capture observed trend and seasonality in monthly collision rates up to around a year ahead.  

Our modelling approach can be improved in a number of ways. For example, it is common to have covariate information such as traffic flow or average speed associated with a particular location at which a collision has occurred. However, pooling such data over zones is time-consuming and not always straightforward. Nevertheless, incorporation of covariates into the DLM framework is straightforward in principle, via the observation equation, and we anticipate improved prediction in this scenario. Although not pursued here, our model can also be used to predict collision rates at zones for which observations are not available. Interpolation of the fitted GP component in the system equation governing the dynamic mean and GP prior over the static parameters governing the harmonic, can be performed for unobserved zones of interest; see e.g. \cite{williams2005} for further details.

\bibliography{References}

\begin{thebibliography}{}

\bibitem[Banerjee et~al., 2014]{banerjee2014hierarchical}
Banerjee, S., Carlin, B.~P., and Gelfand, A.~E. (2014).
\newblock {\em Hierarchical modeling and analysis for spatial data}.
\newblock CRC press.

\bibitem[Buddhavarapu, 2015]{buddhavarapu2015bayesian}
Buddhavarapu, P. N. V. S.~R. (2015).
\newblock {\em On {B}ayesian estimation of spatial and dynamic count models
  using data augmentation techniques: application to road safety management}.
\newblock PhD thesis.

\bibitem[Carter and Kohn, 1994]{carter1994gibbs}
Carter, C.~K. and Kohn, R. (1994).
\newblock On {G}ibbs sampling for state space models.
\newblock {\em Biometrika}, 81(3):541--553.

\bibitem[Fei et~al., 2011]{fei2011bayesian}
Fei, X., Lu, C.-C., and Liu, K. (2011).
\newblock A {B}ayesian dynamic linear model approach for real-time short-term
  freeway travel time prediction.
\newblock {\em Transportation Research Part C: Emerging Technologies},
  19(6):1306--1318.

\bibitem[Fr{\"u}hwirth-Schnatter, 1994]{fruhwirth1994data}
Fr{\"u}hwirth-Schnatter, S. (1994).
\newblock Data augmentation and dynamic linear models.
\newblock {\em Journal of time series analysis}, 15(2):183--202.

\bibitem[Gamerman and Migon, 1993]{gamerman1993dynamic}
Gamerman, D. and Migon, H.~S. (1993).
\newblock Dynamic hierarchical models.
\newblock {\em Journal of the Royal Statistical Society: Series B
  (Methodological)}, 55(3):629--642.

\bibitem[Gilks et~al., 1995]{gilks1995markov}
Gilks, W.~R., Richardson, S., and Spiegelhalter, D. (1995).
\newblock {\em Markov chain {M}onte {C}arlo in practice}.
\newblock CRC press.

\bibitem[Harvey, 1990]{harvey1990forecasting}
Harvey, A.~C. (1990).
\newblock Forecasting, structural time series models and the {K}alman filter.

\bibitem[Lai et~al., 2020]{lai2020sequential}
Lai, Y., Golightly, A., and Boys, R.~J. (2020).
\newblock Sequential {B}ayesian inference for spatio-temporal models of
  temperature and humidity data.
\newblock {\em Journal of Computational Science}, 43:101125.

\bibitem[Petris et~al., 2009]{petris2009dynamic}
Petris, G., Petrone, S., and Campagnoli, P. (2009).
\newblock Dynamic linear models.
\newblock In {\em Dynamic Linear Models with R}, pages 31--84. Springer.

\bibitem[Rasmussen and Williams, 2005]{williams2005}
Rasmussen, C.~E. and Williams, C. K.~I. (2005).
\newblock {\em Gaussian Processes for Machine Learning}.
\newblock The MIT Press.

\bibitem[Roberts and Rosenthal, 2001]{roberts2001optimal}
Roberts, G.~O. and Rosenthal, J.~S. (2001).
\newblock Optimal scaling for various {M}etropolis-{H}astings algorithms.
\newblock {\em Statistical science}, 16(4):351--367.

\bibitem[Shaddick and Wakefield, 2002]{shaddick2002modelling}
Shaddick, G. and Wakefield, J. (2002).
\newblock Modelling daily multivariate pollutant data at multiple sites.
\newblock {\em Journal of the Royal Statistical Society: Series C (Applied
  Statistics)}, 51(3):351--372.

\bibitem[West and Harrison, 2006]{west2006bayesian}
West, M. and Harrison, J. (2006).
\newblock {\em {B}ayesian forecasting and dynamic models}.
\newblock Springer Science \& Business Media.

\bibitem[WHO, 2020]{WHO}
WHO (2020).
\newblock Road traffic injuries.
\newblock
  \url{https://www.who.int/news-room/fact-sheets/detail/road-traffic-injuries}.
\newblock [Online; accessed 24-May-2021].

\end{thebibliography}
\bibliographystyle{apalike}
\end{document}